\documentclass[12pt]{article}
\usepackage{amsfonts}
\begin{document}

\def\ba{\begin{eqnarray}}
\def\ea{\end{eqnarray}}

\title{{\bf A Solution to Symmetric Teleparallel Gravity}}
\author{ Muzaffer ADAK  \\
 {\small Department of Physics, Faculty of Arts and Sciences,}\\
 {\small Pamukkale University,}\\
{\small 20100 Denizli, Turkey} \\ {\small madak@pamukkale.edu.tr} \\ \\
 \"{O}zcan SERT  \\
 {\small Department of Physics, Faculty of Arts and Sciences,}\\
 {\small Pamukkale University,}\\
{\small 20100 Denizli, Turkey} \\ {\small osert@pamukkale.edu.tr}}
\vskip 1cm
\date{\today }
\maketitle

\begin{abstract}

Teleparallel gravity models, in which the curvature and the
nonmetricity of spacetime are both set zero, are widely studied in
the literature. We work a different teleparallel theory, in which
the curvature and the torsion of spacetime are both constrained to
zero, but the nonmetricity is nonzero. After reformulating the
general relativity in this spacetime we find a solution and
investigate its singularity structure.
\end{abstract}

\section{\large Introduction}

Einstein's general relativity provides an elegant
(pseudo-)Riemannian formulation of gravitation in the absence of
matter. In the variational approach, Einstein's field equations
are obtained by considering variations of the Einstein-Hilbert
action with respect to the metric and its associated Levi-Civita
connection of the spacetime. That is, the absence of matter means
that the connection is metric compatible and torsion free, a
situation which is natural but not always convenient. A number of
developments in physics in recent years suggest the possibility
that the treatment of spacetime might involve more than a
Riemannian structure~\cite{muz}.

Theories of gravity based on the geometry of distant parallelism
\cite{hay1}-\cite{arc} are commonly considered as the closest
alternative to the general relativity (GR) theory. Teleparallel
gravity models possess a number of attractive features both from
the geometrical and physical viewpoints. Teleparallelism naturally
arises within the framework of the gauge theory of the group of
general coordinate transformations which underlies GR.
Accordingly, the energy-momentum current represents the matter
source in the field equations of the teleparallel gravity.

Since gauge theories seem important for the description of
fundamental interactions it appears natural to exploit any gauge
structure present in theories of gravity. Different authors,
however, adopt different criteria in order to determine what
properties a theory should possess in order for it to qualify as a
gauge theory. We take the gravitational gauge group to be the
local Lorentz group \cite{ben}.

In this paper we will study a gravity model in a spacetime whose
curvature and torsion are both zero, but the nonmetricity is
nonzero. There is a few work in the literature about gravity
models in this kind of spacetimes; the so-called symmetric
teleparallel gravity~\cite{nes}.

\section{\large Mathematical preliminaries}\label{matpre}

Spacetime is denoted by the triple $ \{M,g,\nabla \} $ where M is
a 4-dimensional differentiable manifold, equipped with a
Lorentzian metric $ g $ which is a second rank, covariant,
symmetric, non-degenerate tensor and $ \nabla $ is a linear
connection which defines parallel transport of vectors (or more
generally tensors and spinors). With an orthonormal basis $\{ X_a
\}$,
 \ba
      g = \eta_{ab}e^a \otimes e^b \;\;\; , \;\; a,b,\cdots = 0,1,2,3
        \label{metric}
 \ea
where $\eta_{ab}=(-,+,+,+)$ is the Minkowski metric and $\{ e^a
\}$ is the orthonormal co-frame. The local orthonormal frame $\{
X_a \}$ is dual to the co-frame $\{e^a \}$;
 \ba
     e^b(X_a)=\delta^b_a \; .
 \ea
The manifold $M$ is oriented with the volume 4-form
 \ba
      { }^*1 = e^0 \wedge e^1 \wedge e^2 \wedge e^3
 \ea
where $*$ denotes the Hodge map and it is convenient to employ in
the following the graded interior operator $\imath_{X_a} \equiv
\imath_a$:
 \ba
     \imath_a e^b = \delta^b_a \; .
 \ea
In addition, the connection $ \nabla $ is specified by a set of
connection 1-forms ${\Lambda^a}_b$. In the  gauge approach to
gravity $\eta_{ab} , \quad e^a , \quad {\Lambda^a}_b$ are
interpreted as the generalized gauge potentials, while the
corresponding field strengths; the nonmetricity 1-forms, torsion
2-forms and curvature 2-forms are defined through the Cartan
structure equations
 \ba
   2Q_{ab} &:=& -D\eta_{ab} = \Lambda_{ab} +\Lambda_{ba} \, ,\label{nonmet}\\
   T^a &:=& De^a = de^a + {\Lambda^a}_b \wedge e^b \, , \label{torsion}\\
  {R^a}_b &:=& D{\Lambda^a}_b :=
     d{\Lambda^a}_b +{\Lambda^a}_c \wedge {\Lambda^c}_b
     \label{curva}
 \ea
where $d$ and $D$ denote the exterior derivative and the covariant
exterior derivative, respectively. These field strengths satisfy
the Bianchi identities\footnote{Since
$Q^{ab}=\frac{1}{2}D\eta^{ab} \neq 0$ we pay special attention in
lowering and raising an index in front of the covariant exterior
derivative.}
 \ba
     DQ_{ab} &=& \frac{1}{2}(R_{ab} +R_{ba}) \label{bian1}\\
     DT^a    &=& {R^a}_b \wedge e^b \label{bian2}\\
     D{R^a}_b &=& 0 \label{bian3} \; .
 \ea
The linear connection 1-forms can be decomposed uniquely as
follows \cite{der},\cite{heh}
 \ba
  {\Lambda^a}_b = {\omega^a}_b + {K^a}_b
          + {q^a}_b + {Q^a}_b \;  \label{connec}
 \ea
where $ {\omega^a}_b $ are the  Levi-Civita connection 1-forms
that satisfy
 \ba
     de^a + {\omega^a}_b \wedge e^b = 0 \; , \label{levi}
 \ea
$ {K^a}_b $ are the contortion 1-forms such that
 \ba
     {K^a}_b \wedge e^b = T^a \; , \label{contor}
 \ea
and $ {q^a}_b $ are the anti-symmetric tensor 1-forms defined by
 \ba
     q_{ab} = -(\imath_a Q_{bc}) \wedge e^c
        + (\imath_b Q_{ac}) \wedge e^c \, . \label{antisy}
 \ea
In the above  decomposition the symmetric part
 \ba
    \Lambda_{(ab)} = Q_{ab} \label{symm}
 \ea
while the anti-symmetric part
 \ba
  \Lambda_{[ab]} = \omega_{ab} + K_{ab} + q_{ab} \; . \label{asymm}
 \ea
It is cumbersome to take into account all components of
nonmetricity in gravitational models. Therefore we will be content
with dealing  only with certain irreducible parts of it to gain
physical insight. The irreducible decompositions of nonmetricity
invariant under the Lorentz group are summarily given below
\cite{heh}. The nonmetricity 1-forms $ Q_{ab} $ can be split into
their trace-free $ \overline{Q}_{ab} $ and the trace parts as
 \ba
   Q_{ab} = \overline{Q}_{ab} + \frac{1}{4} \eta_{ab}Q          \label{118}
 \ea
where the Weyl 1-form $Q={Q^a}_a$ and $ \eta^{ab}\overline{Q}_{ab}
= 0 $. Let us define
 \ba
    \Lambda_b &:=& \imath_a { \overline{Q}^a}_b \;\; , \;\;\;\;\;\;\;\;
                   \Lambda := \Lambda_a e^a,    \nonumber  \\
    \Theta_b &:=& {}^*(\overline{Q}_{ab} \wedge e^a) \;\; , \;\;\;
    \Theta := e^b \wedge \Theta_b \;\; , \;\;\;
    \Omega_a := \Theta_a -\frac{1}{3}\imath_a\Theta    \label{119}
 \ea
as to use them in the decomposition of $ Q_{ab} $ as
 \ba
     Q_{ab} = { }^{(1)}Q_{ab} + { }^{(2)}Q_{ab} +
            { }^{(3)}Q_{ab} + { }^{(4)}Q_{ab}             \label{120}
 \ea
where
 \ba
    { }^{(2)}Q_{ab} &=& \frac{1}{3} {}^*(e_a \wedge \Omega_b +e_b
    \wedge \Omega_a)   \\
    { }^{(3)}Q_{ab} &=& \frac{2}{9}( \Lambda_a e_b +\Lambda_b e_a
                      -\frac{1}{2} \eta_{ab} \Lambda )  \\
    { }^{(4)}Q_{ab} &=&  \frac{1}{4} \eta_{ab} Q    \\
    { }^{(1)}Q_{ab}  &=& Q_{ab}- { }^{(2)}Q_{ab}
                        - { }^{(3)}Q_{ab} - { }^{(4)}Q_{ab} \;.
 \ea
We have $ \imath_a { }^{(1)}Q^{ab} =\imath_a { }^{(2)}Q^{ab} =0 \;
, \quad \eta_{ab} { }^{(1)}Q^{ab} = \eta_{ab} { }^{(2)}Q^{ab}
=\eta_{ab} { }^{(3)}Q^{ab} = 0 $ , \\
 $  e_a \wedge { }^{(1)}Q^{ab} =0 $ and $ \imath_{(a}{ }^{(2)}Q_{bc)} $.

\section{\large Symmetric teleparallel gravity }\label{model}

In the symmetric teleparallel gravity (STPG)~\cite{nes}, we have
two geometrical constraints
 \ba
    {R^a}_b &=& d{\Lambda^a}_b +{\Lambda^a}_c \wedge {\Lambda^c}_b =0 \label{R0}\\
     T^a &=& de^a + {\Lambda^a}_b \wedge e^b =0  \; . \label{T0}
 \ea
These equations mean that there is a distant parallelism, but the
angles and lengths may change during a parallel transport.

In the literature there are many works on teleparallel gravity
models \cite{hay1}-\cite{arc} in which constraints are given
 \ba
    {R^a}_b = 0  \quad , \quad
    {Q^a}_b = 0 \; . \label{R02}
 \ea
One trivial solution to (\ref{R02}) is $\eta_{ab}=(-,+,+,+)$ and
${\Lambda^a}_b=0$. Then the orthonormal co-frame $\{e^a\}$ is left
over as the only dynamical variable. We call such a choice {\it
Weitzenb\"{o}k gauge}. This gauge can not be a solution to STPG
because of equations (\ref{R0}) and (\ref{T0}) since when we set
$\eta_{ab}=(-,+,+,+)$ and ${\Lambda^a}_b=0$ this give rise
identically to $e^a = dx^{\hat{a}}$: the so-called {\it Minkowski
gauge} \cite{nes}.

Now we give a brief outline of GR. GR is written in (pseudo-)
Riemannian spacetime in which torsion and nonmetricity are both
zero, i.e., connection is Levi-Civita. Einstein equation can be
written in the following form
  \ba
     G_a := -\frac{1}{2}R^{bc}(\omega) \wedge { }^*(e_a \wedge e_b \wedge
      e_c)= \kappa \tau_a \label{einsteineqn}
  \ea
or alternative form
 \ba
     { }^*G_a := \mbox{(Ric)}_a - \frac{1}{2} \mathcal{R} e_a = \kappa { }^*\tau_a
 \ea
where $G_a$ is Einstein tensor 3-form, $R^{ab}(\omega)$ is
Riemannian curvature 2-form, $\mbox{(Ric)}_a = \imath_b {R^b}_a
(\omega)$ is Ricci curvature 1-form, $ \mathcal{R} =\imath_a
\mbox{(Ric)}^a $ is scalar curvature, $\tau_a $ is energy-momentum
3-form and $\kappa$ is coupling constant.

For the symmetric teleparallel equivalent of Einstein equation we
first decompose non-Riemannian curvature 2-form (\ref{curva}) via
(\ref{connec}) as follows, with ${K^a}_b =0$
 \ba
    {R^a}_b(\Lambda ) = {R^a}_b(\omega ) + D(\omega )({q^a}_b + {Q^a}_b )
      +({q^a}_c + {Q^a}_c) \wedge ({q^c}_b + {Q^c}_b)
 \ea
where $D(\omega )$ is the covariant exterior derivative with the
Levi-Civita connection. After setting ${R^a}_b(\Lambda ) = 0$ we
obtain the symmetric teleparallel equivalent of
(\ref{einsteineqn})
 \ba
   G_a :=\frac{1}{2} [ D(\omega )q^{bc} + {q^b}_k \wedge q^{kc} + {Q^b}_k \wedge Q^{kc} ]
       \wedge { }^*(e_a \wedge e_b \wedge e_c)  =  \kappa \tau_a \;
       .\label{stpegr}
 \ea

\subsection{Spherical symmetric solution to the
model}\label{solution}

We now proceed the attempt for finding a solution to the STPG
model. As usual in the study of exact solutions, we have two
steps. The first one is to choose the convenient local coordinates
and make corresponding ansatz for the dynamical fields. The second
step concerns providing the invariants of the resulting geometry.
While the choice of an ansatz helps to solve the field equations
easily, the invariant description provides the correct
understanding of the physical contents of a solution.

Since metric and connection are independent quantities in
non-Riemannian spacetimes, we have to predict separately
appropriate candidates for them. Therefore we first write a line
element in order to determine the metric. We naturally start
dealing with the case of spherical symmetry for realistic
simplicity,
 \ba
     g=-F^2 dt^2 + G^2dr^2 + r^2d\theta^2 +r^2\sin^2\theta
     d\varphi^2
 \ea
where $F=F(r)$ and $G=G(r)$. A convenient choice for a tetrad
reads
 \ba
      e^0 = Fdt , \quad e^1= Gdr , \quad e^2= rd\theta ,
      \quad e^3 = r\sin\theta d\varphi \; . \label{coframe}
 \ea
In addition, for the non-Riemannian connection we choose
 \ba
     \Lambda_{12} &=& -\Lambda_{21}= - \frac{1}{r}e^2 , \quad
     \Lambda_{13}=-\Lambda_{31}= -  \frac{1}{r}e^3  , \quad
     \Lambda_{23}=-\Lambda_{32}= - \frac{\cot\theta}{r}e^3 , \nonumber \\
     \Lambda_{00} &=& \frac{F'}{FG}e^1 , \quad
     \Lambda_{11} = \frac{1}{r}(1-\frac{1}{G})e^1 , \quad
     \Lambda_{22} = \frac{1}{r}(1-\frac{1}{G})e^1 , \nonumber \\
     \Lambda_{33} &=& \frac{1}{r}(1-\frac{1}{G})e^1 , \quad \mbox{others}=0  \label{connect} \; .
 \ea
These gauge configurations (\ref{coframe}) and (\ref{connect})
satisfy the constraint equations ${R^a}_b(\Lambda )=0 \; , \quad
T^a(\Lambda )=0$. One can certainly perform a locally Lorentz
transformation
 \ba
      e^a \rightarrow {L^a}_b e^b \quad , \quad
      {\Lambda^a}_b \rightarrow {L^a}_c {\Lambda^c}_d {{L^{-1}}^d}_b
      +{L^a}_c d{{L^{-1}}^c}_b
 \ea
which yields the Minkowski gauge ${\Lambda^a}_b =0$. This may mean
that we propose a set of connection components in a special frame
and coordinate which seems contrary to the spirit of relativity
theory. However in physically natural situations we can choose a
reference and coordinate system at our best convenience.

We deduce from equations (\ref{coframe})-(\ref{connect})
 \ba
    \omega_{01} &=& -\frac{F'}{FG}e^0 , \quad \omega_{12}=- \frac{1}{rG} e^2 , \quad
     \omega_{13}=- \frac{1}{rG} e^3 , \quad
     \omega_{23}= -\frac{\cot\theta}{r} e^3 \nonumber \\
  Q_{00} &=& \frac{F'}{FG}e^1 , \quad Q_{11} = \frac{1}{r}(1-\frac{1}{G})e^1 , \quad
     Q_{22} = \frac{1}{r}(1-\frac{1}{G})e^1 , \quad
     Q_{33} = \frac{1}{r}(1-\frac{1}{G})e^1  \nonumber \\
   q_{01} &=& \frac{F'}{FG}e^0 , \quad q_{12} = \frac{1}{r}(\frac{1}{G}-1)e^2 , \quad
     q_{13} =\frac{1}{r}(\frac{1}{G}-1) e^3  , \quad   \mbox{others}=0 \; .
     \label{oQq}
 \ea
When we put (\ref{oQq}) into (\ref{stpegr}) we obtain, with
$\tau_a = 0 $
 \ba
     \left( dq^{bc} + 2{\omega^b}_f \wedge q^{fc}
      + {q^b}_f \wedge q^{fc} \right) \wedge { }^*(e_a \wedge e_b \wedge e_c) =0
 \ea
whose components read explicitly
 \ba
     Zeroth \;\; component \quad \quad \quad \quad  \quad \quad \;\;
                                 \left[ \frac{2(G^{-1})'}{rG} - \frac{G^2 -1}{r^2G^2} \right]
                                        e^1 \wedge e^2 \wedge e^3 = 0 \label{zeroth}\\
     First \;\; component \quad \quad \quad \quad  \quad \quad \quad
                                        -\left[ \frac{2F'}{rFG^2} - \frac{G^2 -1}{r^2G^2} \right]
                                        e^0 \wedge e^2 \wedge e^3 = 0 \label{first}\\
     Second \;\; component \quad \quad \;\; \left[ \frac{(F'G^{-1})'}{FG}
                                    + \frac{F'}{rFG^2} + \frac{(G^{-1})'}{rG} \right]
                                       e^0 \wedge e^1 \wedge e^3 = 0 \label{second}\\
     Third \;\; component \quad \quad - \left[ \frac{(F'G^{-1})'}{FG} + \frac{F'}{rFG^2} + \frac{(G^{-1})'}{rG} \right]
                                        e^0 \wedge e^1 \wedge e^2 =
                                        0 \; . \label{third}
 \ea
Then from (\ref{zeroth}) and (\ref{first})
 \ba
       G(r) &=& 1/F(r)
 \ea
and from (\ref{second}) and (\ref{third})
 \ba
      F^2(r)&=& 1- \frac{C}{r}
 \ea
where $C$ is a constant.

In order to have a correct understanding of the resulting
solution, we need to construct invariants of the {\it Riemannian}
curvature and nonmetricity. Although the total curvature is
identically zero in the teleparallel gravity, the Riemannian
curvature of the Levi-Civita connection is nontrivial:
 \ba
   R^{01}(\omega ) =\frac{(F'G^{-1})'}{FG} e^{10} \quad , \quad
   R^{02}(\omega ) =\frac{F'}{rFG^2} e^{20} \quad , \quad
   R^{03}(\omega ) =\frac{F'}{rFG^2} e^{30} \quad , \nonumber \\
   R^{12}(\omega ) =\frac{(G^{-1})'}{rG} e^{21} \quad , \quad
   R^{13}(\omega ) =\frac{(G^{-1})'}{rG} e^{31} \quad , \quad
   R^{23}(\omega ) =\frac{1}{r^2}(1-\frac{1}{G^2}) e^{32} \; .
 \ea
Thus the quadratic invariant of the Riemannian curvature reads
 \ba
  R_{ab}(\omega ) \wedge { }^*R^{ab}(\omega ) &=&
                         \left\{ 2  \left[ \frac{(F'G^{-1})'}{FG}
                         \right]^2 \right.
                     + 4 \left( \frac{F'}{rFG^2} \right)^2
                         + 4 \left[ \frac{(G^{-1})'}{rG} \right]^2
                     +  \left. 2 \left[ \frac{1}{r^2} \left( 1-\frac{1}{G^2} \right) \right]^2 \right\} {
                     }^*1 \nonumber \\
  &=& \frac{6C^2}{r^6} { }^*1 \label{rinvar}
 \ea
and the spacetime geometry is naturally characterized by the
quadratic invariant of the nonmetricity
 \ba
     Q_{ab}\wedge { }^*Q^{ab} &=&
       \left\{ \left( \frac{F'}{FG} \right)^2
          + 3 \left[ \frac{1}{r} \left( 1-\frac{1}{G} \right) \right]^2
          \right\} { }^*1 \nonumber \\
       &=& \left\{ \frac{C^2}{4r^3(r-C)} - \frac{3C}{r^3} + \frac{6}{r^2}
        \left[ 1- \left( 1- \frac{C}{r}\right)^{1/2} \right] \right\} { }^*1 \; .
        \label{qinvar}
 \ea
These two quadratic invariants provide the sufficient tools for
understanding the contents of the classical solutions.

\section{Discussion}

In this paper we have studied a gravity model in the spacetime
only with nonmetricty. A similar analysis, the so-called symmetric
teleparallel gravity (STPG), was performed in~\cite{nes}. The main
motivation for studying STPG is to determine the place and
significance of the symmetric teleparallel GR-equivalent model
since the GR-equivalent models are satisfactorily supported by
observations. Thus we hope to gain physical insights to
nonmetricity. Important observation is that the Riemannian
curvature invariant (\ref{rinvar}) is singular at $r=0$, but
regular at the zero ($r=C$) of the metric function $F(r)$, which
means that we have a horizon here. The resulting geometry then
describes the well known Schwarzschild black hole at $r=0$ with
the horizon at $r=C$. Since we are dealing with symmetric
teleparallel gravity, it is necessary also to analyze the behavior
of nonmetricity. As seen from (\ref{qinvar}), the nonmetricity
invariant diverges not only at the origin $r=0$, but also at the
Schwarzschild horizon $r=C$. The horizon is a regular surface from
the viewpoint of the Riemannian geometry, but it is singular from
the viewpoint of symmetric teleparallel gravity. We intend to
clarify the geometrical and physical meaning of the singularities
in STPG by investigating matter coupling to STPG in a separate
paper.

 \vskip 1.5cm

\noindent {\large {\bf Acknowledgement}}

\noindent This work is supported by the Scientific Research
Project (BAP) 2002FEF007, Pamukkale University, Denizli, Turkey.


\end{document}